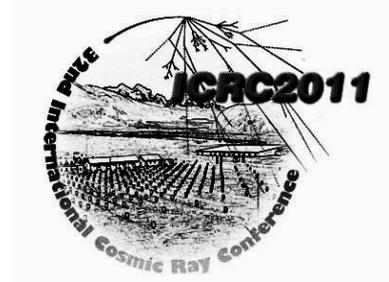

# The UFFO (Ultra Fast Flash Observatory) Pathfinder: Science and Mission

P. CHEN[1], S. AHMAD[2], K. AHN[3], P. BARRILLON[2], S. BLIN-BONDIL[2], S. BRANDT[4], C. BUDTZ-JØRGENSEN[4], A.J. CASTRO-TIRADO[5], H.S. CHOI[6], Y.J. CHOI[7], P. CONNELL[8], S. DAGORET-CAMPAGNE[2], C. DE LA TAILLE[2], C. EYLES[8], B. GROSSAN[9], I. HERMANN[7], M.-H. A. HUANG[10], S. JEONG[11], A. JUNG[11], J.E. KIM[11], S.H. KIM[3], Y.W. KIM[11], J. LEE[11], H. LIM[11], E.V. LINDER[9,11], T.-C. LIU[1], NIELS LUND[4], K.W. MIN[7], G.W. NA[11], J.W. NAM[11], K. NAM[11], M.I. PANAYUK[12], I.H. PARK[11], V. REGLERO[8], J.M. RODRIGO[8], G.F. SMOOT[9,11], Y.D. SUH[7], S. SVELITOV[12], N. VEDENKEN[12], M.-Z WANG[1], I. YASHIN[12], M.H. ZHAO[11] [THE UFFO COLLABORATION]

[1]*National Taiwan University, Taipei, Taiwan*
[2]*University of Paris-Sud 11, France*
[3]*Yonsei University, Seoul, Korea*
[4]*National Space Institute, Denmark*
[5]*Instituto de Astrofisica de Andalucia, Consejo Superior de Investigaciones Cientificas, Spain*
[6]*Korea Institute of Industrial Technology, Ansan, Korea*
[7]*Korea Advanced Institute of Science and Technology, Daejeon, Korea*
[8]*University of Valencia, Spain*
[9]*University of California, Berkeley, USA*
[10]*National United University, Miao-Li, Taiwan*
[11]*Ewha Womans University, Seoul, Korea*
[12]*Moscow State University, Moscow, Russia*

pisinchen@phys.ntu.edu.tw; chen@slac.stanford.edu

**Abstract:** Hundreds of gamma-ray burst (GRB) optical light curves have been measured since the discovery of optical afterglows. However, even after nearly 7 years of operation of the *Swift* Observatory, only a handful of measurements have been made soon (within a minute) after the gamma ray signal. This lack of early observations fails to address burst physics at short time scales associated with prompt emissions and progenitors. Because of this lack of sub-minute data, the characteristics of the rise phase of optical light curve of short-hard type GRB and rapid-rising GRB, which may account for ~30% of all GRB, remain practically unknown. We have developed methods for reaching sub-minute and sub-second timescales in a small spacecraft observatory. Rather than slewing the entire spacecraft to aim the optical instrument at the GRB position, we use rapidly moving mirror to redirect our optical beam. As a first step, we employ motorized slewing mirror telescope (SMT), which can point to the event within 1s, in the UFFO Pathfinder GRB Telescope onboard the *Lomonosov* satellite to be launched in Nov. 2011. UFFO's sub-minute measurements of the optical emission of dozens of GRB each year will result in a more rigorous test of current internal shock models, probe the extremes of bulk Lorentz factors, provide the first early and detailed measurements of fast-rise GRB optical light curves, and help verify the prospect of GRB as a new standard candle. We will describe the science and the mission of the current UFFO Pathfinder project, and our plan of a full-scale UFFO-100 as the next step.

**Keywords:** Gamma Ray Burst,

## 1 Importance of GRB Prompt Signal

Much progress has been made in Gamma Ray Burst (GRB) science since the launch of the Swift observatory [1] 7 years ago. The observations from Swift did not, however, produce a simple picture of GRB, but rather documented the richness and complexity of this phenomenon. Just a few years ago, GRBs were believed to be of only two types, distinguished only by their gamma-X emission: a shorter, hard spectrum burst, with duration of gamma-X emission less than two seconds, and a longer, soft burst. After 476 observations by *Swift* Burst Alert Telescope (BAT) made between Dec. 2004 and Dec. 2009 (BAT2 Catalog) [2] and its follow-up UV-optical observation by the *Swift* UV-Optical Telescope (UVOT), a huge variation in optical light curves has been observed, especially in the early rise time. Figure 1 shows a sample of the optical light curve measurements made soon after GRB triggers. There appear to be distinct classes of fast-rising ($t_{peak} < 10^2$ s) and slow-rising bursts [1].

Additionally, the optical light curves are complex, with decays, plateaus, changes in slope, and other features that are not yet understood. Panaitescu and Vestrand [3] claim that the optical luminosity distribution of the fast-rising bursts at ~$10^3$ s is quite narrow, and has promise as a kind of "standard candle" which would make GRBs useful as a cosmological probe of the very high redshift universe. In order to move this possible trend to the status of a refined tool, a larger sample of such objects is required, and in particular, better resolution is required at early times. Are there more features in the early light curve that are missed by the sparse sampling? Does any feature of the rise correlate with the luminosity or a particular aspect of the physics? How many bursts are misclassified because the rapid rise was missed? The need for earlier measurements (faster optical response after the initial gamma-ray burst) is clear and compelling.

The challenge is particularly acute for short-hard GRB observations, which have few early measurements. What is the shape of the rise? Is the shape homogeneous? The physical origin of this type of burst remains an outstanding mystery, so any hint as to its origin would be extremely valuable. Because of the short time scale for the gamma-X light curves and the lower bolometric luminosity, these bursts are believed to originate from the merger of compact objects. Is there any prompt UV-optical emission from such events? What would we see if we observed more of these events in the sub-minute or sub-second regime? Are there ultra-short events on the accretion disk dynamical timescale of compact objects (that are beamed so we can see them)? Early observations would answer these questions and open a new window probing compact object structure, populations, and evolution.

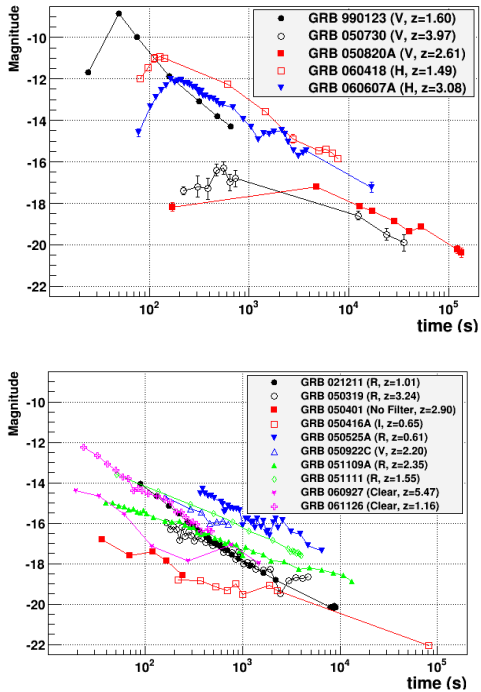

Figure 1. GRB optical light curve rise time and shape [3] for the Fast-Rising class (top) and the Decay class (bottom).

In GRB 080319B, extraordinary, bright, variable optical emission, which peaked at the visual magnitude of 5.3, has been observed while the prompt gamma-ray emission was still active. This observation clearly shows that there can exist a prompt optical emission component that tracks the gamma-ray light curve [4]. This further motivates us to push for the observation of optical emissions in the GRB prompt signals.

In addition to providing the first early and detailed measurements of fast-rise GRB optical light curves and helping verify the prospect of GRB as a new standard candle, other science potentials from systematic observations of GRB prompt signals include a more rigorous test of current internal shock models and the probing of the extremes of bulk Lorentz factors.

## 2 A New Approach: Steering the Optical Path, not the Spacecraft

In the *Swift* observatory, the entire spacecraft slews to point its UV-Optical Telescope (UVOT) at the GRB position after the Burst Alert Telescope (BAT) [7] identifies the onset of the event. The histogram of burst event as a function of response time falls off below 100 s, with an almost complete cutoff at 60 s. Due to its finite mission lifetime, *Swift* cannot be expected to significantly increase the number of sub-minute response events.

To circumvent this challenge, our approach is to redirect the optical path of the incoming GRB beam instead of the entire spacecraft. One exciting prospect is to invoke a micro electrical mechanical system (MEMS) to build a mirror array (MMA), where each pixel of mirror is controlled by a nano-fabricated micro-motor and therefore no massive mechanical motion is required. Such a MMA can redirect the telescope beam to a target within one msec. The technology is reasonably mature. The Research Center for MEMS Space Telescope (RCMST) at Ewha Women's University, Korea, a key institution in the UFFO Collaboration, has successfully invoked this approach in two recent satellite missions, KAMTEL and MTEL [5]. The challenge, however, is the uniformity of the MEMS array for large area mirrors. While further R&D is required, we believe that this would become the future paradigm for flash or transient observation telescopes.

## 3 UFFO-Pathfinder

With the aforementioned science motivations in mind, the Taiwan and Korea members of the present UFFO Collaboration initiated the POET (Prompt Observation of Energetic Transients) satellite project in 2008, but it was later aborted. Afterwards, the UFFO-Pathfinder was proposed in 2009 in response to an opportunity afforded by available space aboard the *Lomonosov* Space Mission, scheduled for launch in Nov. 2011. UFFO is a multinational project that involves scientists from Denmark, France, Korea, Russia, Spain, Taiwan, and the U.S. [6]. The *Lomonosov* UNIVSERSAT spacecraft will carry several instruments, including the TUS air shower cos-mic ray experiment. Through collaboration with Moscow State University, the UFFO Pathfinder proposal was granted for X-ray triggered observations of GRB through an optical telescope.

UFFO-Pathfinder consists of two major components: the Slewing Mirror Telescope (SMT) and the UFFO Burst Alert and Trigger Telescope (UBAT). We briefly describe their characteristics below.

### 3.1 Slewing Mirror Telescope

Because of the time constraint for meeting the launching schedule, the UFFO Collaboration has decided not to employ the MEMS mirror array but a semi-conventional approach for the UFFO Pathfinder slewing mirror telescope (SMT). It is a Ritchey-Chrétien telescope with motorized gimbal-mounted mirror 10 cm in diameter, with a field of view of 17×17 arcmin. Other specifications of SMT are given in Table 1. Under the Lomonosov mass constraint, we roughly evenly divide the allotted total mass of 21.5 kg to UFFO's two key components: SMT and UBAT. With only 11.5 kg in weight, SMT can redirect the incoming GRB light path within 1s.

| Telescope | Ritchey-Chrétien + motorized mirror plate |
|---|---|
| Aperture | 10 cm diameter |
| F-number | 11.4 |
| Detector and / Operation | Intensified CCD with MCP/ Photon Counting |
| Field of View | 17 x 17 arcmin |
| Detection Element | 256 x 256 pixels |
| Telescope PSF | 1 arcsec @ 350 nm |
| Pixel Scale | 4 arcsec |
| Wavelength Range | 200 nm – 650 nm |
| Sensitivity | 17.5 mag/10 s exposure 18.7 mag/100 s exposure (5 sigma, open filter) |
| Bright Limit | mv = 6 mag |
| Data taking start time after trigger+location | < 1 s |
| Data Rate | 1 GB/day |
| Mass, Power consumption, Size | 11.5 kg, 10W, 30cm (W) x 20cm (H) x 62cm (L) |

Table 1. Specifications of UFFO Pathfinder SMT.

### 3.2 UFFO Burst Alert and Trigger Telescope

The UFFO Burst Alert & Trigger telescope (UBAT) will be similar to the Swift BAT X-gamma trigger camera, using a coded mask aperture camera scheme for good position detection for transients and wide field of view.

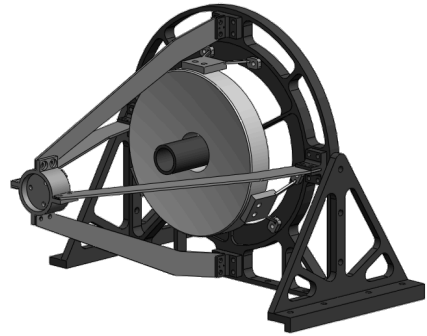

Figure 2. A rendering of the opto-mechanics of the UFFO Pathfinder slewing mirror telescope (SMT).

However, in order to respond over a wider energy range, making the camera more sensitive to broad-band hard sources including GRB, a design including the LYSO crystal and 64 (8x8) MAPMT as the detector is invoked, resulting in a sensitive energy range of 5-200 keV. With a mass constraint of 10 kg for UBAT, we use a detection area of 191 cm$^2$. The resulting sensitivity is 310 mCrab in 10 s at 5 σ. The specifications of UBAT are given in Table 2 and a 3D view of it is shown in figure 3. Figure 4 shows the integrated UFFO system.

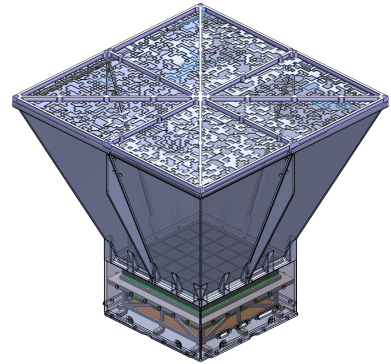

Figure 3. A rendering of the UFFO Burst Alert and Trigger Telescope (UBAT)

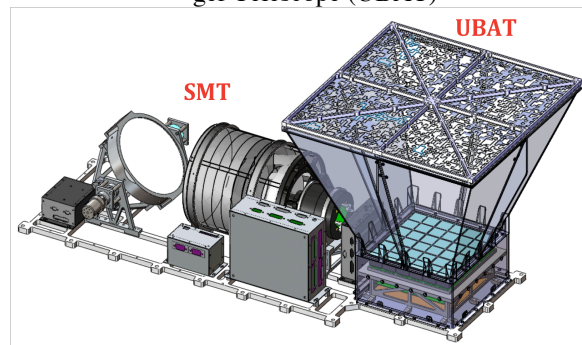

Figure 4. The integrated UFFO Pathfinder. The SMT enclosure is not shown in this drawing.

## 4 Expected Results and Impact

In order to estimate our event rate, we examined the fluence distribution of Swift BAT GRB that triggered *Swift* UVOT observations during the first 7 years of

| Overall | Mass of the camera | 10 kg |
|---|---|---|
| | Energy range | 5 – 200 keV |
| | Telescope PSF | ≤ 17 arcmin |
| | Source position accuracy | ≤ 10 arcmin (> 7σ) |
| | Field of view | ~1.85 sr (90.2° x 90.2°) |
| | GRB detection rate | ~ 43/yr |
| Detector plane | Compounds | LYSO + MAPMT |
| | Effective area | 191 cm$^2$ |
| | Pixel size | 2.88 x 2.88 x 2 mm$^3$ |
| | Number of pixels | 48 x 48 |
| | Spectral energy Resolution | 20% at 60 keV |
| | Sensitivity | 310 mCrab for 10 sec exposures at 5σ, 5-100 keV |
| Passive shielding | Compounds (out to in) | Al, W |
| | Absorption @ 4-50 keV | 100 % |
| Coded mask | Compounds | W alloy |
| | Total size | 392 x 392 mm$^2$ |
| | Mask to detector plane distance | 28 cm |

Table 2. Key parameters of UFFO Burst Alert and Trigger Telescope (UBAT).

*Swift* operation. This is a more conservative number than the total rate of *Swift* BAT GRB. Scaling by our estimated sensitivity to that of *Swift* BAT, we find that we will still receive an expected ~43 GRB triggers for SMT per year from UBAT. Of these, we expect ~2 short-hard triggers per year. The actual number of SMT observations that we accomplish should cover about the same number, unless our orbit has significantly more restrictions than that of Swift. It is likely that some reduction in these numbers could result from our inability to point away from the galactic plane, unlike Swift. Because short bursts are hard, and because our detectors have more low-energy response than BAT, some reduction in the short-hard rate may result.

Barring significant malfunctions, UFFO will provide sub-minute UV-optical measurements for dozens of GRB observations within the first year of operation, with about 9 detections. These optical measurements will be the first ever under 10 s after the gamma ray trigger, and will make up the first ever large-sample, systematic survey of optical emission in the sub-minute regime.

We note that there is an exciting synergy with the TUS instrument. In the case that a GRB produces neutrino or other cosmic ray signals, extensive air showers are expected. Our experiment would detect the source GRB event in X-gamma and UV-optical photons, while the TUS instrument would detect the particle shower, measuring the source position and arrival times of the particles with 0.1° and 10 $\mu$s accuracy, respectively. Such measurements would enable the first measurements of neutrino masses from GRB emission, and would serve as an exciting new measure of photon and particle dispersion relations, of great interest to fundamental particle physics, cosmology and relativity tests.

## 5 Future Prospect: UFFO-100

While still busy with the preparation of UFFO-Pathfinder, the UFFO Collaboration has been exploring its next step, a more ambitious project: UFFO-100, based on the same design principle but with larger total mass of 100 kg (thus the name UFFO-100). This would afford a 30 cm aperture slewing telescope and a 1024 cm$^2$ CZT X-ray camera. The goal is to finally integrate the MMA technology with the motorized slewing mirror and to add an IR-sensitive camera to detect the distinguished bursts. The key components and the dimensions of UFFO-100 are shown in figure 5. We expect to launch UFFO-100 in 2015 by the Soyuz Launcher.

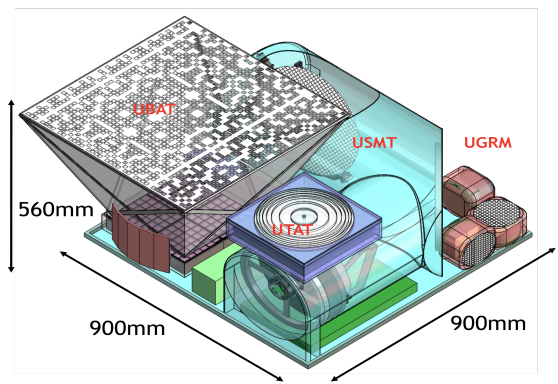

Figure 5. A rendering of the UFFO-100 GRB Telescope

## 6 Summary

The UFFO Pathfinder has now entered the final stage of preparation before it is launched in Nov. 2011. We eagerly look forward to its exciting GRB findings and the proof-of-principle for this new approach to future GRB telescopes.